arXiv:0907.1966

# A nonextensive approach for the instability of current-driven ion-acoustic waves in space plasma

Liu Zhipeng, Liu Liyan, Du Jiulin*

*Department of Physics, School of Science, Tianjin University, Tianjin 300072, China*

**Abstract**

The instability of current-driven ion-acoustic waves in the collisionless magnetic-field-free space plasma is investigated by using a nonextensive approach. The ions and the electrons are thought of in the power-law distributions that can be described by the generalized $q$-Maxwellian velocity distribution and are considered with the different nonextensive $q$-parameters. The generalized q-wave frequency and the generalized instability $q$-growth rate for the ion-acoustic waves are derived. The numerical results show that the nonextensive effects on the ion-acoustic waves are not apparent when the electron temperature is much more than the ion temperature, but they are salient when the electron temperature is *not* much more than the ion temperature. As compared with the electrons, the ions play a dominant role in the nonextensive effects.

## I. INTRODUCTION

Ion-acoustic waves in plasma belong to the class of longitudinal electrostatic waves and exhibit strongly nonlinear properties. This kind of waves can be observed commonly in space and laboratory plasmas. The theoretical and experimental results show that the ion-acoustic waves play an important role in the turbulence heating, the laser plasma interaction, the particle acceleration and other various aspects etc. When examining the investigations on the current-driven ion-acoustic instability, one can find that nearly all the works have employed the Maxwellian velocity distribution as the statistical description for the stationary distribution of the ions and electrons. However, computer simulations revealed that the electron distribution was not purely

* E-mail address: jiulindu@yahoo.com.cn

Maxwellian [1]. In some physical situations of either laboratory plasma or space plasma, the velocity distributions are not Maxwellian one but behave the power-law. For example, the spacecraft measurements for the plasma velocity distribution indicated that there was a "suprathermal" power-law tail at the high energies [2], which can be modeled more effectively by a generalized Lorentzian κ-distribution [3]. Such a κ–distribution is known now to be equal to the *q*-distribution in nonextensive statistics [4,5].

In this paper, we apply nonextensive statistics to study the current-driven ion-acoustic wave instability in space plasmas. From a new point of view, we give one theoretical explanation about the high energy tail observed in space plasmas [6]. We assume that the ions and the electrons have the generalized *q*-Maxwellian velocity distribution with the different nonextensive *q*-parameters. Their nonextensive effects on the waves and instability are analyzed respectively.

In 1988, Tsallis proposed the nonextensive entropy or the q-entropy [7]. Since then the statistics based on the *q*-entropy has been developed and now called nonextensive statistics. It is becoming a useful approach to the statistical property for various complex systems. The *q*-entropy can be written by

$$S_q = k_B \frac{\int (f^q - f) d^3x d^3v}{1-q}, \tag{1}$$

where $f$ is the probabilistic distribution function, the parameter $q$ is a real number different from unity, specifying the degree of nonextensivity. Boltzmann-Gibbs entropy is obtained from $S_q$ if taking the limit $q \to 1$. An important property of the q-entropy is the nonadditivity or nonextensivity for $q \neq 1$. For example, for one system composed of two parts $A$ and $B$, the total q-entropy of the composite system $A+B$ is

$$S_q(A+B) = S_q(A) + S_q(B) + k_B^{-1}(1-q) S_q(A) S_q(B). \tag{2}$$

Nonextensive statistics has been applied to the long-range interacting systems, among which self-gravitating systems and plasma systems might offer the best framework for searching into the nonextensive effects because the long-range interactions between particles play a fundamental role in determining the properties of such systems [8]. For instance, the cosmic microwave background radiation is studied using the power-law q-distribution [9], thermonuclear reactions rates in stellar plasmas are calculated based on nonextensive statistics [10]. The nonextensivity in the nonequilibrium plasma with Coulombian long-range interactions is investigated using the generalized q-Maxwellian velocity distribution [11], in which the formulation for the nonextensive parameter $q$ is found and the physical meaning is given. Nonextensive statistics is also applied to study the waves and instability phenomena, such as the plasma oscillations [12,13], the relativistic Langmuir waves [14], the linear or nonlinear Landau damping [15] and, recently, Jeans' instability [16,17].

The organization of this paper is as follows: in section II, we study the generalized dispersion relation and the generalized growth rate of instability as described by nonextensive statistics. Then, in section Ⅲ, we analyze the nonextensive effects on the wave frequency and the growth rate. The illustrative figures are given. Finally, the conclusion is given in section Ⅳ.

## II. THE GENERALIZED DISPERSION RELATION AND THE GROWTH RATE OF INSTABILITY

In nonextensive statistics, the generalized $q$-Maxwellian velocity distribution function can be written by the form [11,18],

$$f_0(v) = \frac{n_0 A_q}{\sqrt{\pi} v_T}\left[1-(q-1)\frac{v^2}{v_T{}^2}\right]^{1/(q-1)} \tag{3}$$

$$A_q = \sqrt{1-q}\,\frac{\Gamma(\frac{1}{1-q})}{\Gamma(\frac{1}{1-q}-\frac{1}{2})} \text{ for } 0<q\le 1, \text{ and } A_q = \frac{1+q}{2}\sqrt{q-1}\,\frac{\Gamma(\frac{1}{1-q}+\frac{1}{2})}{\Gamma(\frac{1}{q-1})} \text{ for } q\ge 1,$$

where $v_T$ is thermal speed, $v_T = \sqrt{2k_B T/m}$, $n_0$ is particle number density, $T$ is temperature, and $m$ is particle mass. The Maxwellian velocity distribution can be recovered from Eq.(3) in the limit $q \to 1$. For $q > 1$, there is a thermal cutoff for this distribution, $v_c = v_T\sqrt{1/(q-1)}.$

One assumes the plasma to be collisionless and magnetic-field-free and at the q-equilibrium with the distribution Eq.(3). When the plasma departs slightly from the q-equilibrium, the distribution function can be approximately written as

$$f_\alpha = f_{\alpha 0}(v) + f_{\alpha 1}(\vec{v},\vec{r},t), \quad f_{\alpha 1} \ll f_{\alpha 0} \tag{4}$$

where the subscript $\alpha$ refers to particle species, with $\alpha = i, e$ for ions and electrons, respectively, $f_{\alpha 0}$ is the power-law q-equilibrium distribution of species $\alpha$ and $f_{\alpha 1}$ is the perturbation about $f_{\alpha 0}$. If the electron component has drift relative to the ion component, the *q*-distribution functions for the electrons and the ions are written, respectively, as

$$f_{e0}(v) = \frac{n_{e0}}{\sqrt{\pi}} \frac{A_{q_e}}{v_{Te}} \left[ 1 - (q_e - 1) \frac{(v - V_\mathrm{d})^2}{v_{Te}^2} \right]^{1/(q_e - 1)} \tag{5}$$

$$f_{i0}(v) = \frac{n_{i0}}{\sqrt{\pi}} \frac{A_{q_i}}{v_{Ti}} \left[ 1 - (q_i - 1) \frac{v^2}{v_{Ti}^2} \right]^{1/(q_i - 1)}, \tag{6}$$

where $q_e$ and $q_i$ are the nonextensive parameters for the electrons and the ions, respectively, $V_d$ is the velocity of the electrons relative to the ions, and $v_{T\alpha}$ is the thermal speed. Usually, one assumes the plasma to be in the low-drift-velocity region, $V_d \ll v_{Te} = (2k_B T_e / m_e)^{1/2}$. Let both the static electric field and the magnetic field be zero, $\mathbf{E_0} = \mathbf{B_0} = 0$, and the perturbed magnetic field be zero, $\mathbf{B_1} = 0$. Then the Vlasov's and Poisson's equation for such a system are given, respectively, by

$$\frac{\partial f_\alpha}{\partial t} + \mathbf{v} \cdot \nabla f_\alpha - \frac{Q_\alpha}{m} \nabla \phi \cdot \nabla_\mathrm{v} f_\alpha = 0 \tag{7}$$

$$\nabla^2 \phi = \sum_\alpha \nabla^2 \phi_\alpha = -\frac{1}{\varepsilon_0} \sum_\alpha Q_\alpha \int f_{\alpha 1} d\mathbf{v}, \tag{8}$$

where $\phi$ is the electrostatic potential, satisfying $\mathbf{E_1} = -\nabla \phi$, $\mathbf{E_1}$ is the electric field intensity produced by the perturbation, $Q_\alpha$ is the charge of the species $\alpha$. Without lose of generality, the x-axis is along the direction of wave vector $\mathbf{k}$, $v_x = u$. Taking the perturbations in the form of $f_{\alpha 1}, \phi \propto \exp[i(kx - \omega t)]$, and making Fourier transformation for Eq. (7) and (8), we can lead to the equation,

$$\left[ 1 - \sum_\alpha \frac{\omega_{p\alpha}^2}{k^2} \frac{1}{n_{\alpha 0}} \int \frac{\partial f_{\alpha 0} / \partial u}{(u - \omega / k)} du \right] \phi = 0, \tag{9}$$

where $\omega_{p\alpha} = \sqrt{n_{\alpha 0} e^2 / \varepsilon_0 m_\alpha}$ is called the natural oscillation plasma frequency. Since the electrostatic potential has non-trivial solution, the term in the brackets must be zero, which gives the dispersion relation,

$$\varepsilon_q\left(\omega,k\right)=1-\sum_{\alpha}\frac{\omega_{p\alpha}^2}{k^2}\frac{1}{n_{\alpha 0}}\int\frac{\partial f_{\alpha 0}/\partial u}{\left(u-\omega/k\right)}du=0\,, \tag{10}$$

where $\varepsilon_q\left(\omega,k\right)$ is the dielectric permittivity. Eq.(10) is a familiar form in textbook [19]. It holds for any equilibrium distribution function [20]. Now substituting the $q$-distributions Eq.(5) and (6) into Eq.(10), after some algebra, one derives the generalized $q$-dispersion relation of the current-driven ion-acoustic waves,

$$1+\frac{1}{k^2\lambda_{De}^2}\left[\frac{1+q_e}{2}+(\xi_e-V_d')Z_{q_e}(\xi_e-V_d')\right]+\frac{1}{k^2\lambda_{Di}^2}\left[\frac{1+q_i}{2}+\xi_i Z_{q_i}(\xi_i)\right]=0\,, \tag{11}$$

where $\lambda_{D\alpha}=v_{T\alpha}/\sqrt{2}\omega_{p\alpha}$ is the Debye length of the species $\alpha$, $\xi_\alpha$ is a dimensionless parameter defined as the ratio of the phase velocity, $v_\phi=\omega/k$, to the thermal speed $v_{T\alpha}=\sqrt{2k_BT_\alpha/m_\alpha}$, namely, $\xi_\alpha=v_\phi/v_{T\alpha}$, and $V_d'=V_d/v_{Te}$ is the rate between the speeds. $Z_{q_\alpha}(\xi_\alpha)$ is called the generalized $q$-dispersion function in nonextensive statistics [20],

$$Z_{q_\alpha}(\xi_\alpha)=\frac{A_{q_\alpha}}{\sqrt{\pi}}\int_{-\infty}^{+\infty}\frac{\left[1-\left(q_\alpha-1\right)x^2\right]^{(2-q_\alpha)/(q_\alpha-1)}}{x-\xi_\alpha}dx\,. \tag{12}$$

Obviously, in the limit $q_\alpha\to 1$, Eq.(12) recovers the standard Fried-Conte dispersion function,

$$Z(\xi)=\frac{1}{\sqrt{\pi}}\int_{-\infty}^{\infty}\frac{e^{-x^2}}{x-\xi}dx\,. \tag{13}$$

If the ion temperature is much lower than the electron temperature and the ion mass is much heavier than the electron mass, one has $v_{Ti}\ll v_\phi\ll v_{Te}$ and, from the dimensionless parameter $\xi_\alpha=v_\phi/v_{T\alpha}$, one gets $\xi_e\ll 1$ and $\xi_i\gg 1$. Making the asymptotic expansion of Eq.(12) for the ion and the electron, respectively, and then integrating it, one obtains

$$1+\frac{1}{k^2\lambda_{De}^2}\left\{\frac{1+q_e}{2}+i\sqrt{\pi}(\xi_e-V_d')A_{q_e}\left[1-\left(q_e-1\right)(\xi_e-V_d')^2\right]^{\frac{2-q_e}{q_e-1}}\right\}$$
$$+\frac{1}{k^2\lambda_{Di}^2}\left\{\left(\frac{1}{2\xi_i^2}+\frac{3}{2\left(3q_i-1\right)}\right)+i\sqrt{\pi}\xi_iA_{q_i}\left[1-\left(q_i-1\right)\xi_i^2\right]^{\frac{2-q_i}{q_i-1}}\right\}=0.$$

(14)

The dielectric permittivity $\varepsilon_q\left(\omega,k\right)$ can be written in terms of the real and imaginary parts as $\varepsilon_q\left(\omega,k\right)=\varepsilon_q^r\left(\omega,k\right)+i\varepsilon_q^i\left(\omega,k\right)$. If the high-order terms are neglected, then $\varepsilon_q^r$ and $\varepsilon_q^i$ can be written approximately as

$$\varepsilon_q^r \approx 1+\frac{1}{k^2\lambda_{De}^2}\frac{1+q_e}{2}+\frac{1}{k^2\lambda_{Di}^2}\left[-\frac{1}{2\xi_i^2}-\frac{3}{2\left(3q_i-1\right)}\frac{1}{\xi_i^4}\right] \tag{15}$$

$$\begin{aligned}\varepsilon_q^i &= \frac{\sqrt{\pi}}{k^2\lambda_{De}^2}A_{q_e}(\xi_e-V_d')\left\{\left[1-\left(q_e-1\right)\left(\xi_e-V_d'\right)^2\right]\right\}^{\frac{2-q_e}{q_e-1}}+\frac{\sqrt{\pi}}{k^2\lambda_{Di}^2}A_{q_i}\xi_i\left[1-\left(q_i-1\right)\xi_i^2\right]^{\frac{2-q_i}{q_i-1}}\\ &\approx \frac{\sqrt{\pi}}{k^2\lambda_{De}^2}A_{q_e}(\xi_e-V_d')+\frac{\sqrt{\pi}}{k^2\lambda_{Di}^2}A_{q_i}\xi_i\left[1-\left(q_i-1\right)\xi_i^2\right]^{\frac{2-q_i}{q_i-1}}.\end{aligned} \tag{16}$$

We denote $\omega=\omega_r+i\gamma_q$, where $\omega_r=\mathrm{Re}\,\omega$ is the oscillation frequency and $\gamma_q=\mathrm{Im}\,\omega$ is the growth rate of the waves. When $\gamma_q>0$, the wave amplitude grows with the increase of time, corresponding to the instability. Let the real part of the dielectric permittivity to be zero, $\varepsilon_q^r\left(\omega_r,k\right)=0$, then we get the generalized $q$-dispersion relation,

$$\frac{\omega_r}{\omega_{pe}}=k\lambda_{De}\sqrt{\frac{m_e}{m_i}}\left[\frac{1}{\left(1+q_e\right)/2+k^2\lambda_{De}^2}+\frac{6}{3q_i-1}\frac{T_i}{T_e}\right]^{1/2}. \tag{17}$$

This equation is related to two different nonextensiv parameters, $q_e$ for the electron and $q_i$ for the ion. If we take $q_e=q_i=q$, Eq.(17) reduces to the dispersion relation obtained by Liu and Du [20],

$$\frac{\omega_r^2}{k^2}=\frac{1}{\left(1+q\right)/2+k^2\lambda_{De}^2}\frac{k_BT_e}{m_i}+\frac{3}{3q-1}v_{T_i}^2. \tag{18}$$

If in the limit, $q_e\to 1, q_i\to 1$, Eq.(17) recovers to the plasma frequency in the case of using Maxwellian velocity distribution,

$$\frac{\omega_r^2}{k^2}=\frac{1}{1+k^2\lambda_{De}^2}\frac{k_BT_e}{m_i}+\frac{3}{2}v_{T_i}^2 . \tag{19}$$

For $\gamma_q \ll \omega_r$, we may make the small-argument expansion [19] for the dielectric function,

$$\begin{aligned}\varepsilon_q(\omega,k) &= \varepsilon_q^r(\omega_r+i\gamma_q,k)+i\varepsilon_q^i(\omega_r+i\gamma_q,k)\\ &\approx \varepsilon_q^r(\omega_r,k)+i\gamma_q\frac{\partial\varepsilon_q^r}{\partial\omega_r}\bigg|_{\omega=\omega_r}+i\varepsilon_q^i(\omega_r,k)=0.\end{aligned} \tag{20}$$

Then the instability growth rate can be written as

$$\gamma_q=-\frac{\varepsilon_q^i(\omega_r,k)}{\partial\varepsilon_q^r/\partial\omega\big|_{\omega=\omega_r}}. \tag{21}$$

Using Eq.(15) and (16), the generalized growth rate can be obtained form Eq.(21) as

$$\begin{aligned}\frac{\gamma_q}{\omega_{pe}} =& -\sqrt{\frac{\pi}{8}}\left(\frac{2}{1+q_e+2k^2\lambda_{De}^2}+\frac{6T_i/T_e}{3q_i-1}\right)^2\left(\frac{m_e}{m_i}\right)k\lambda_{De}\left\{A_{q_e}\left[1-V_d'\left(\frac{m_i}{m_e}\right)^{\frac{1}{2}}\left(\frac{4}{1+q_e+2k^2\lambda_{De}^2}+\frac{12T_i/T_e}{3q_i-1}\right)^{-\frac{1}{2}}\right]\right.\\ &\left.+A_{q_i}\left(\frac{T_e}{T_i}\right)^{\frac{3}{2}}\left(\frac{m_i}{m_e}\right)^{\frac{1}{2}}\left[1-(q_i-1)\left(\frac{T_e/T_i}{1+q_e+2k^2\lambda_{De}^2}+\frac{3}{3q_i-1}\right)\right]^{\frac{2-q_i}{q_i-1}}\right\}.\end{aligned} \tag{22}$$

The instability takes place if $\gamma_q$>0, which is equivalent to $V_d$> $V_d^{th}$ (the threshold electron drift velocity). Let the terms in the large brackets in Eq.(22) be zero, we find

$$\begin{aligned}V_d^{th} =& v_{Te}\left(\frac{m_e}{m_i}\right)^{\frac{1}{2}}\left(\frac{4}{1+q_e+2k^2\lambda_{De}^2}+\frac{12}{3q_i-1}\frac{T_i}{T_e}\right)^{\frac{1}{2}}\\ &\times\left\{1+\frac{A_{q_i}}{A_{q_e}}\left(\frac{T_e}{T_i}\right)^{\frac{3}{2}}\left(\frac{m_i}{m_e}\right)^{\frac{1}{2}}\left[1-(q_i-1)\left(\frac{T_e/T_i}{1+q_e+2k^2\lambda_{De}^2}+\frac{3}{3q_i-1}\right)\right]^{\frac{2-q_i}{q_i-1}}\right\}.\end{aligned} \tag{23}$$

If let $V_d=0$ and $q_e=q_i=q$, one can recover the ion-acoustic Landau Damping previously with the Tsallis distribution [20],

$$\begin{aligned}\frac{\gamma_q}{\omega_r} =& -\sqrt{\frac{\pi}{8}}\left[\frac{1}{(1+q)/2+k^2\lambda_{De}^2}+\frac{6}{3q-1}\frac{T_i}{T_e}\right]^{3/2}\\ &\times\left\{\sqrt{\frac{m_e}{m_i}}+\left(\frac{T_e}{T_i}\right)^{3/2}\left[1-(q-1)\left(\frac{T_e/T_i}{1+q+2k^2\lambda_{De}^2}+\frac{3}{3q-1}\right)\right]^{(2-q)/(q-1)}\right\}.\end{aligned} \tag{24}$$

If taking $q = 1$, Eq.(24) reduces to the expression of $\gamma$ in standard Maxwellian distribution,

$$\frac{\gamma}{\omega_{pe}} = -\sqrt{\frac{\pi}{8}}\left(\frac{1}{1+k^2\lambda_{De}^2}+3\frac{T_i}{T_e}\right)^2\left\{1-V_d'\left(\frac{m_i}{m_e}\right)^{\frac{1}{2}}\left(\frac{2}{1+k^2\lambda_{De}^2}+6\frac{T_i}{T_e}\right)^{-\frac{1}{2}}\right.$$
$$\left.+\left(\frac{T_e}{T_i}\right)^{\frac{3}{2}}\left(\frac{m_i}{m_e}\right)^{\frac{1}{2}}\left[\exp\left(\frac{T_e/T_i}{2+2k^2\lambda_{De}^2}+\frac{3}{2}\right)\right]\right\}. \tag{25}$$

## III. NUMERICAL RESULTS AND DISCUSSIONS

In this section, by using the numerical calculations we illustrate the dependence of the wave frequency and the instability growth rate on selected values of the $q_i$-parameter and the $q_e$-parameter.

Fig.1(a) and Fig.1(b) are the numerical curves coming from the q-dispersion relation, Eq.(17), where the normalized wave frequency, $\omega_r/\omega_{pe}$, as a function of the dimensionless parameter, $k\lambda_{De}$, is plotted with the different values of $q_e$ and $q_i$. Here we give the curves for two different cases:

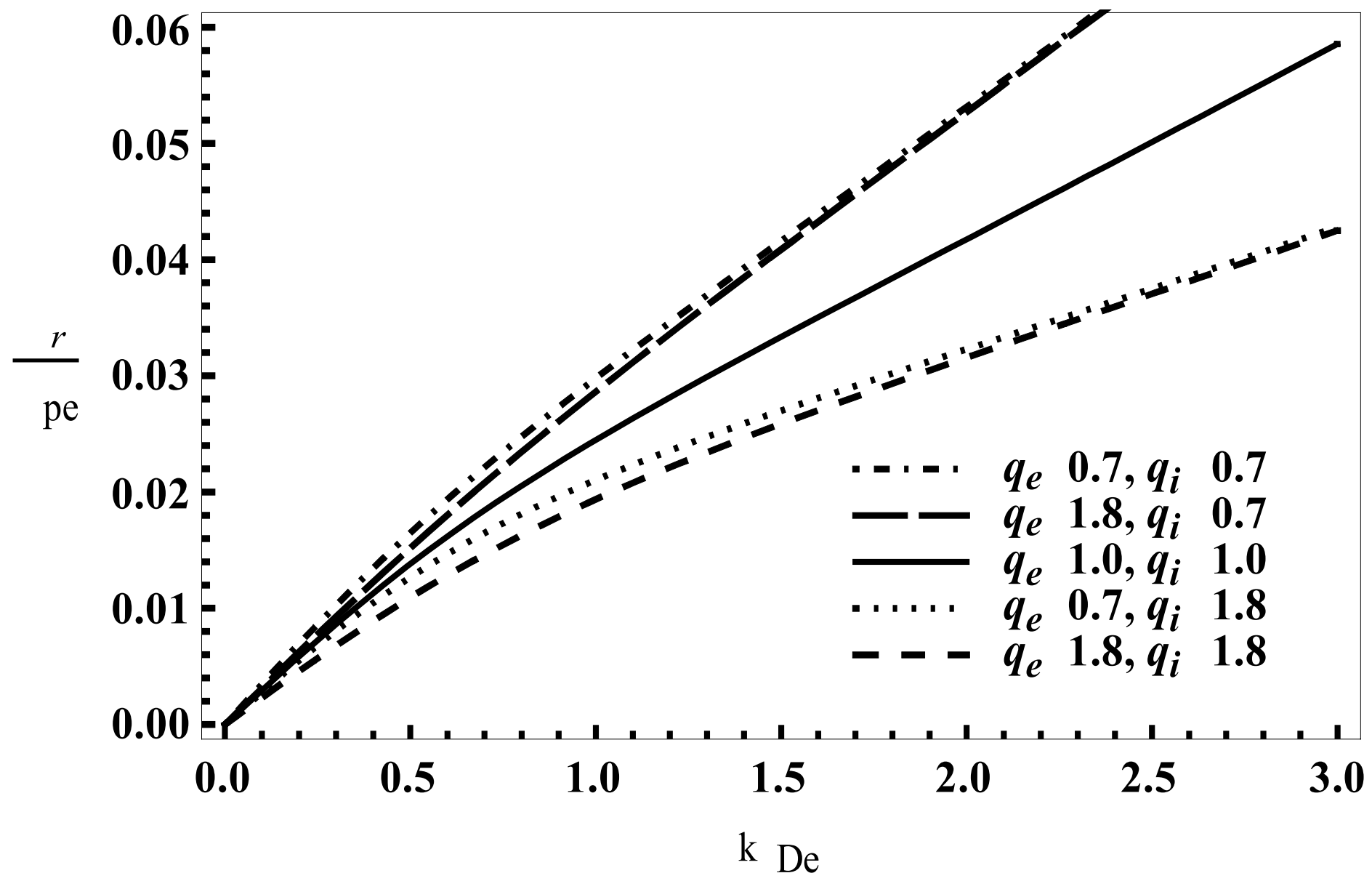

Fig.1 (a) $\omega_r/\omega_{pe}$ as a function of $k\lambda_{De}$ for the different values of $q_i$ and $q_e$ in the case of $T_i/T_e = 1/5$

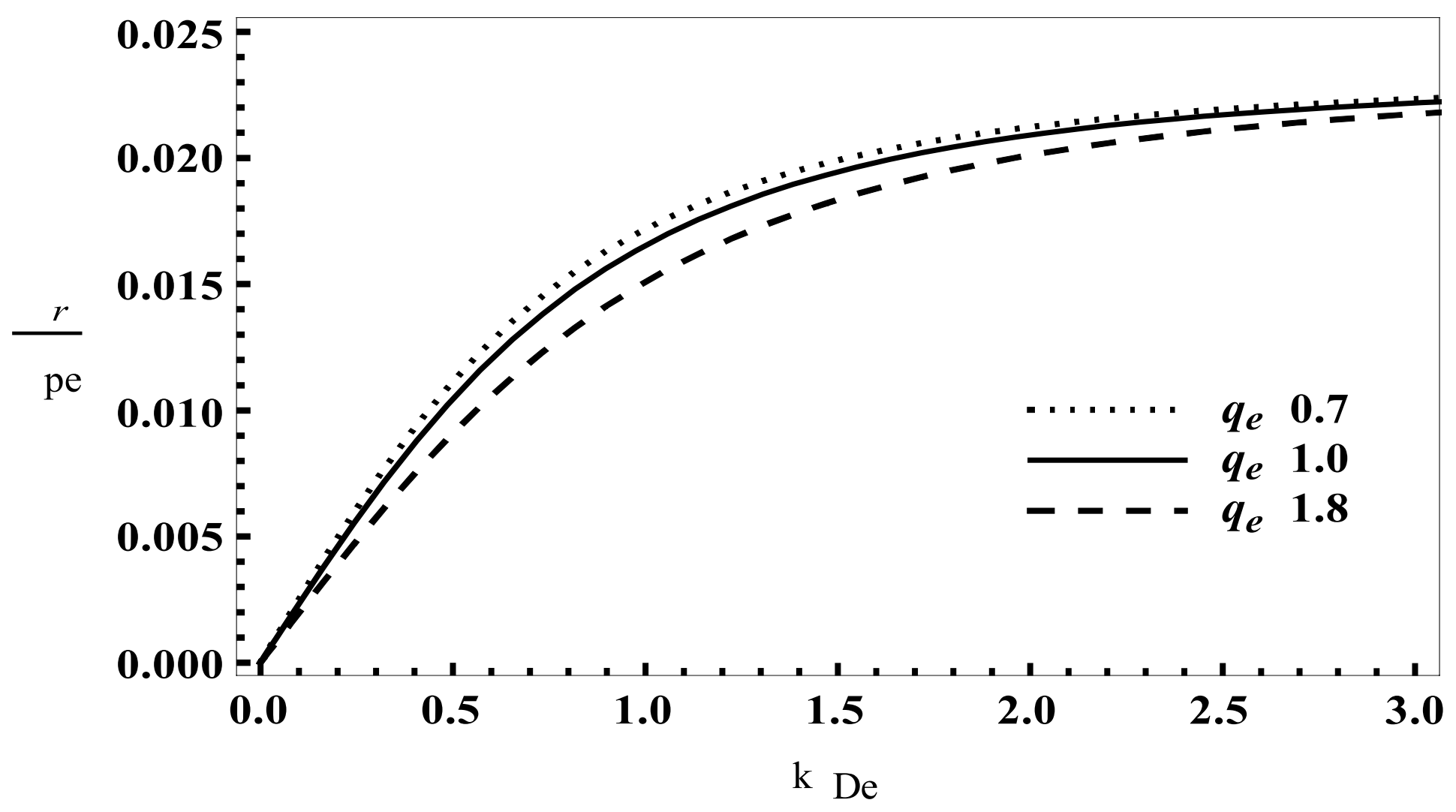

Fig.1 (b) $\omega_r / \omega_{pe}$ as a function of $k\lambda_{De}$ for the different values of $q_i$ and $q_e$ in the case of $T_e$ □ $T_i$.

Fig.1(a) is the curves in the case of $T_e$ %o$T_i$ (i.e. $T_i / T_e = 1/5$). The two curves with $q_i = 1.8$, while $q_e = 0.7$ and 1.8, respectively, are almost superposition; the two curves with $q_i = 0.7$ while $q_e = 0.7$ and 1.8, respectively, are also almost superposition. But they all have a distinct difference from the curve with $q_e = q_i = 1$ for the standard Maxwellian one. It is shown that the wave frequency is affected by the parameter $q_i$ much more than by the parameter $q_e$.

Fig.1(b) is the curves in the case of $T_e$ □ $T_i$, which is quite different from the case (a). In this case, the second term in the large bracket of Eq.(17) can be neglected, so the parameter $q_i$ has little effect on the frequency. Because the parameter $q_i$ has little effect on the frequency, the curves with different values of $q_e$ are almost superposition to the standard Maxwellian one. It is shown that the nonextensive effect is not salient and the increase of the frequency with the growing of $k\lambda_{De}$ slows up, as compared with the case (a).

Fig.2(a) and Fig.2(b) are the numerical curves coming from the generalized growth rate, Eq.(22), where the generalized growth rate $\gamma_q / \omega_{pe}$ as a function of $k\lambda_{De}$ is plotted with the different values of $q_e$ and $q_i$. Taking $V_d / v_{Te} = 1/3$ in Eq.(22), we give the curves for two different cases:

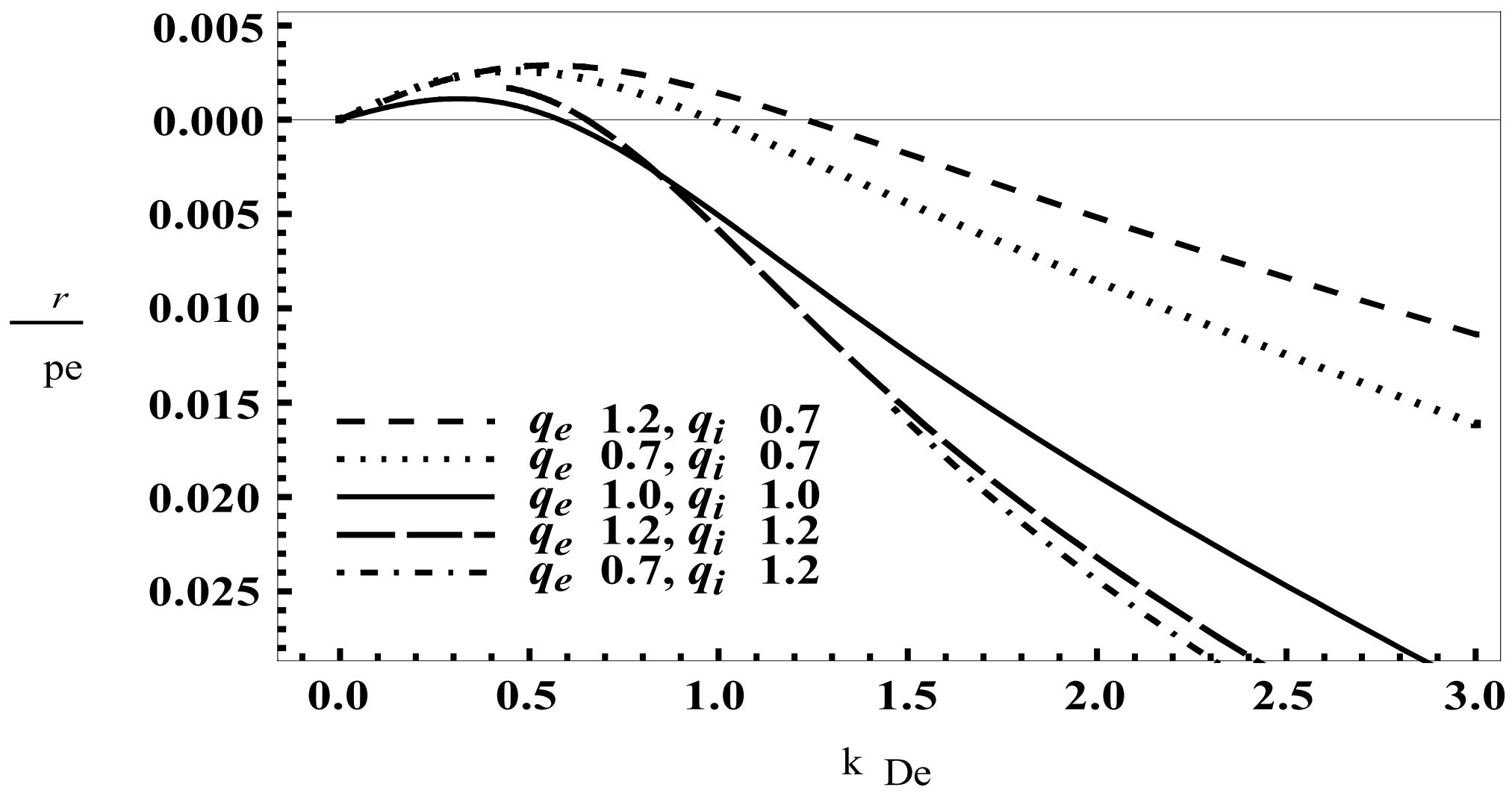

Fig.2.(a) the generalized growth rate $\gamma_q / \omega_{pe}$ is as a function of $k\lambda_{De}$ in the case of $T_i / T_e = 1/5$

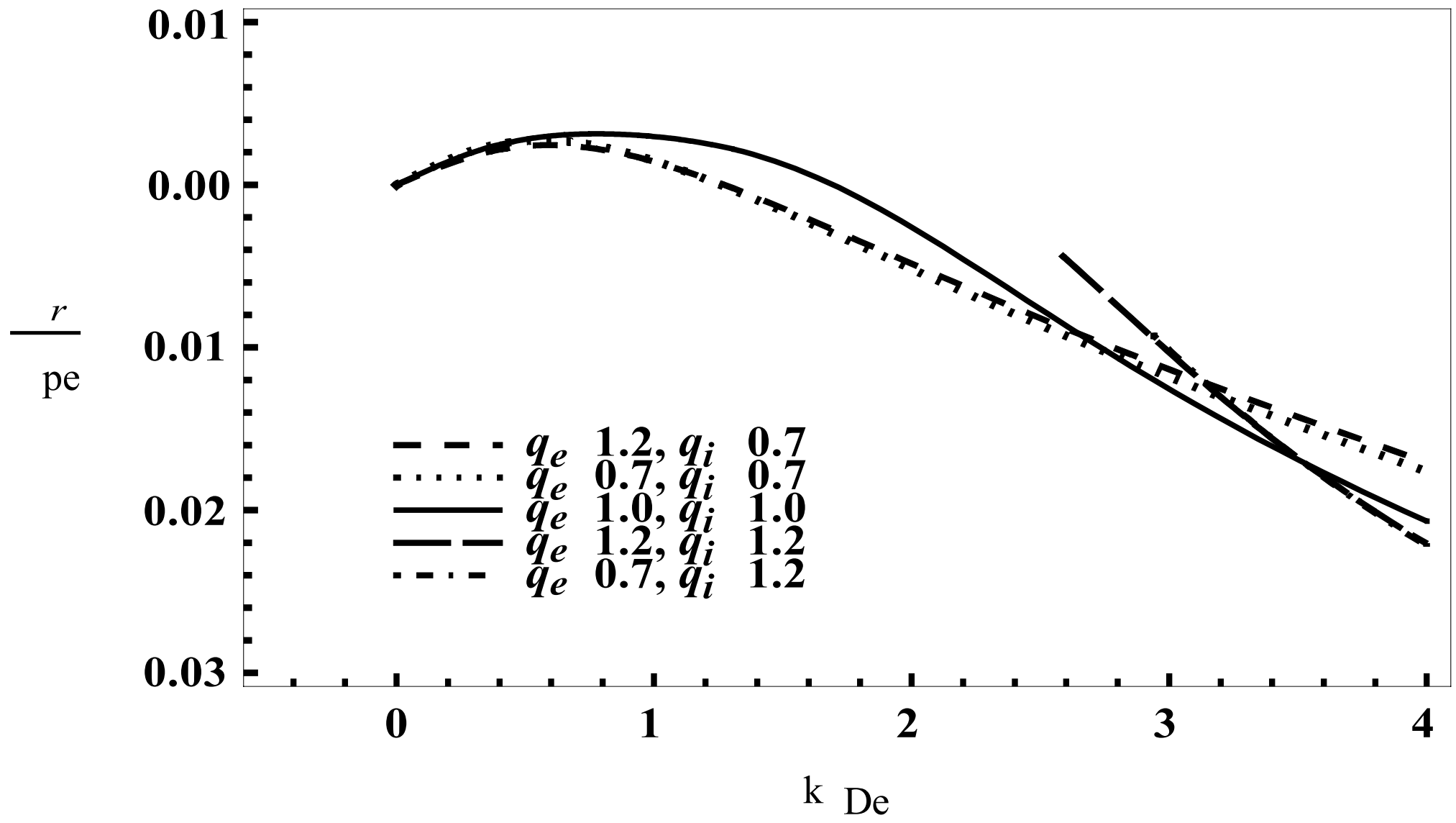

Fig.2. (b) the generalized growth rate $\gamma_q / \omega_{pe}$ is as a function of $k\lambda_{De}$ in the case of

$T_i / T_e = 1/30$

Fig.2(a) is the curves in the case of $T_e$ ‰$T_i$ (i.e. $T_i / T_e = 1/5$ ). It is shown that the generalized growth rate declines rapidly with the increase of the parameter $q_i$ , while the parameter $q_e$ has less effect on the growth rate as compared with $q_i$ . They all have a distinct difference from the curve with $q_i = q_e = 1$ for the standard Maxwellian one. The nonextensive effect is salient.

Fig.2.(b) is the curves in the case of $T_e \gg T_i$ (i.e. $T_i / T_e = 1/30$ ). The ion-electron temperature ratio is very small. In this case, we find that the nonextensive effect on wave growth is not apparent.

## IV. CONCLUSIONS

In conclusion, we employed a nonextensive approach to study the current-driven ion-acoustic instability in some space plasmas. When considering that the electrons and the ions have different nonextensive parameters, $q_e$ and $q_i$, respectively, we have derived the generalized *q*-dispersion relation, Eq.(17), and the generalized instability *q*-growth rate, Eq.(22), for the plasma as described by nonextensive statistics. We find that the two nonextensive parameters, $q_e$ and $q_i$, play different roles in Eq.(17) and Eq.(22). Only if taking $q_e = q_i = q$, are Eq.(17) and Eq.(22) recovered to the forms obtained previously with the Tsallis distribution.[20] The numerical results show that the nonextensive effects are unapparent when the electron temperature is much more than the ion temperature, but the nonextensive effects are salient when the electron temperature is *not* much more than the ion temperature. In each case for the ion-electron temperature ratio, the parameter $q_i$ plays a dominant role as compared with the parameter $q_e$ in the nonextensive effects on the plasma.

Finally, we pay attention to the nonextensive parameters, $q_e$ and $q_i$, for the electrons and the ions. The nonextensive parameter is of crucial importance for nonextensive statistics and the applications, the physical meanings of which were interpreted theoretically by the formulations, [8, 11] and were investigated experimentally by helioseismological measurements.[21] Although the formulations were presented in terms of the Coulombian potential [11] or the gravitational potential,[8] yet the potential can be actually any one.. The two parameters can be calculated directly when one applies the new *q*-dispersion relation and the new instability *q*-growth rate to space plasma. On the basis of the formulation of the nonextensive parameter found for the nonequilibrium plasma,[11] the nonextensive parameters $q_\alpha \,(\alpha = i, e)$ can be written in terms of the relation,

$$k_B \nabla T_\alpha + (1 - q_\alpha) Q_\alpha \nabla \phi = 0, \tag{26}$$

where $Q_\alpha = -e$ for $\alpha = e$ and $Q_\alpha = Ze$ for $\alpha = i$; the potential function, $\phi = \sum_{\alpha=i,e} \phi_\alpha$, is determined by the Poisson's equation, Eq.(8). With these expressions, the two nonextensive parameters both have the clear physical meanings in the plasma.

**ACKNOWLEDGMENT**

We would like to acknowledge the National Natural Science Foundation of China under grant No. 10675088 for the financial supports.